\documentclass[12pt,authoryear]{article}
\usepackage{amsfonts, graphicx,color,multirow, amsthm,amsmath,natbib, hypernat}
\def\boxit#1{\vbox{\hrule\hbox{\vrule\kern6pt\vbox{\kern6pt#1\kern6pt}\kern6pt\vrule}\hrule}}

\newtheorem{theorem}{\indent \sc Theorem}

\newcommand{\argmax}[0]{\mbox{argmax}}
\newcommand{\argmin}[0]{\mbox{argmin}}

\newcommand{\bb}{\mbox{\bf b}}

\newcommand{\bA}{\mbox{\bf A}}
\newcommand{\ba}{\mbox{\bf a}}

\newcommand{\bD}{\mbox{\bf D}}

\newcommand{\bI}{\mbox{\bf I}}

\newcommand{\bS}{\mbox{\bf S}}
\newcommand{\bs}{\mbox{\bf s}}
\newcommand{\bT}{\mbox{\bf T}}
\newcommand{\bX}{\mbox{\bf X}}

\newcommand{\bY}{\mbox{\bf Y}}
\newcommand{\bZ}{\mbox{\bf Z}}

\newcommand{\bbP}{\mathbb{P} }

\newcommand{\bzero}{\mbox{\bf 0}}

\newcommand{\bpsi}{\mbox{\boldmath $\psi$}}

\newcommand{\bbeta}{\mbox{\boldmath $\beta$}}
\newcommand{\btheta}{\mbox{\boldmath $\theta$}}

\newcommand{\hbbeta}{\widehat{\bbeta}}
\newcommand{\hbpsi}{\widehat{\bpsi}}
\newcommand{\hbtheta}{\widehat{\btheta}}

\newcommand{\bgamma}{\mbox{\boldmath $\gamma$}}

\newcommand{\cov}{\mbox{cov}}
\newcommand{\beq}{\begin{eqnarray}}
\newcommand{\eeq}{\end{eqnarray}}
\newcommand{\beqn}{\begin{eqnarray*}}
\newcommand{\eeqn}{\end{eqnarray*}}

\newcommand{\sgn}{\mbox{sgn}}

\begin{document}
\title{\bf Penalized Q-Learning for Dynamic Treatment Regimes
\thanks{Rui Song is Assistant Professor, Department of Statistics, Colorado State University, Fort Collins, CO 80523 (Email: song@stat.colostate.edu). Weiwei Wang is Assistant Professor, Biostatistics/Epidemiology/Research Design (BERD) Core Center for Clinical and Translational Sciences, The University of Texas Health Science Center at Houston, Houston, TX 77030 (Email: Weiwei.Wang@uth.tmc.edu). Donglin Zeng is Associate Professor, Department of Biostatistics, University of North Carolina at Chapel Hill, Chapel Hill, NC 27599 (Email: dzeng@bios.unc.edu). Michael R. Kosorok is Professor and Chair, Department of Biostatistics, University of North Carolina at Chapel Hill, Chapel Hill, NC 27599 (Email: kosorok@unc.edu). We thank the STAR*D team for providing the data for our illustration. STAR*D was supported from the National Institute of Mental Health. We thank Dr. Bibhas Chakraborty for sharing programming codes. Rui Song's research was supported in part by the National Science Foundation grant DMS-1007698. Donglin Zeng's and Michael R Kosorok's research was supported in part by National Institute of Health grant CA142538.}}

\author{Rui Song, Weiwei Wang, Donglin Zeng and Michael R. Kosorok}
\date{May 13, 2010}
\maketitle

\begin{center}
Summary
\end{center}
A dynamic treatment regime effectively incorporates both accrued information and long-term effects of treatment from specially designed clinical trials. As these become more and more popular in conjunction with longitudinal data from clinical studies, the development of statistical inference for optimal dynamic treatment regimes is a high priority. This is very challenging due to the difficulties arising form non-regularities in the treatment effect parameters. In this paper, we propose a new reinforcement learning framework called penalized Q-learning (PQ-learning), under which the non-regularities can be resolved and valid statistical inference established. We also propose a new statistical procedure---individual selection---and corresponding methods for incorporating individual selection within
PQ-learning. Extensive numerical studies are presented which compare the proposed methods with existing methods, under a variety of non-regular scenarios, and demonstrate that the proposed approach is both inferentially and computationally superior. The proposed method is demonstrated with the data from a depression clinical trial study.


\section{Introduction}
Developing effective therapeutic regimens for diseases is one of the
essential goals of medical research. Two major design and analysis
challenges in this effort are: taking accrued information into account
in clinical trial designs and effectively incorporating long-term
benefits and risks of treatment due to delayed effects. One of the
most promising approaches to deal with these two challenges has been
recently referred to as ``dynamic treatment regimes'' or ``adaptive
treatment strategies'' (Murphy, 2003), and the method has been
utilized in a number of settings, such as drug and alcohol dependency studies.

Reinforcement learning---one of the primary tools used in developing dynamic treatment regimes--- is a sub-area of machine learning, where the learning behavior is through trial-and-error interactions with a dynamic environment \citep{KMA96}. Because reinforcement learning techniques have been shown to be effective in developing optimal dynamic treatment regimes, the area is attracting increased attention among statistical researchers. As a recent example, a new approach to cancer clinical trials based on the specific area of reinforcement learning called Q-learning, has been proposed by \cite{ZKZ09}. Extensive statistical estimating methods have also been proposed for optimal dynamic treatment regimes, including, for example, \cite{Cha09}, who developed a Q-learning framework based on linear models. Other related literature includes likelihood-based methods (both frequentist and Bayesian) by \cite{8Thall00,13Thall02,14Thall07} and semiparametric methods by \cite{10Murphy03,Robins04,15Lunceford02}, \cite{9Wahed04,16Wahed06}, \cite{Moodie09} and \cite{MoodieTo}.

In contrast to the substantial body of estimating methods, the development of statistical inference for optimal dynamic treatment regimes is very limited and far from ready. This sequential, multi-stage decision making problem is at the intersection of machine learning, optimization and statistical inference and is thus quite challenging.  As discussed in \cite{Robins04}, and recognized by many other researchers, the key difficulty lies in the fact that the treatment effect parameters at any stage prior to the last stage may be non-regular for certain longitudinal distributions of the data, where non-regularity in this instance means that the asymptotic distribution of the estimator of the treatment effect parameter does not converge uniformly over the parameter space \citep{Cha09}.

This non-regularity arises when the optimal last stage treatment is non-unique for at least some subjects in the population, causing estimation bias and failure of traditional inferential approaches.  There have been a number of proposals for correcting this problem. For example, Moodie and Richardson (2010) proposed a method called Zeroing Instead of Plugging In (ZIPI). This method is also referred to as the hard-threshold estimator by \cite{Cha09}. \cite{Cha09} also proposed a soft-threshold estimator and implemented several kinds of bootstrap methods. Both the hard-threshold estimator and the soft-threshold estimator essentially shrink the ``problematic'' term to decrease the degree of non-regularity. While this intuitively makes sense, there is, however, a lack of theoretical support for these methods. Moreover, extensive simulation studies in their associated papers indicate that neither hard-thresholding nor soft-thresholding, in conjunction with their bootstrap implementation, works uniformly well for all simulation settings. We are therefore motivated to develop improved, asymptotically valid estimation and inference for optimal dynamic treatment regimes.

In this paper, we develop a new reinforcement learning framework for discovering optimal dynamic treatment regimes: Penalized Q-learning (abbreviated hereafter as PQ-learning). This new backward recursive multistage learning approach can be viewed as a penalized version of Q-learning. The major distinction of the proposed PQ-learning from traditional Q-learning is in the form of the objective Q-function at each stage. While the proposed method shares many of the properties of traditional Q-learning, there are at least three significant advantages which we now describe.

First, the notorious and inevitable non-regularity issue associated with Q-learning can be resolved with PQ-learning. At each stage of Q-learning, the maximization functional over individual treatments are involved, hence there is at least one nondifferentiable point over the range of the treatment parameters. If the probability mass on this point is positive, i.e., some individuals have no treatment effects, it will cause non-ignorable non-regularity issues that will yield failure of existing inferential methods. With PQ-learning, all individuals experiencing no treatment effect can be identified with probability converging to one, as in the oracle setting.

Second, we propose effective inferential procedures based on PQ-learning for optimal dynamic treatment regimes. In contrast to existing bootstrap approaches, our variance calculations are based on explicit formula and hence are much less time-consuming. Thorough theoretical studies and extensive empirical evidence both support the validity of the proposed methods.

% which one initiate which? individual selection to PQ-learning? or PQ-learning for individual selection?
Third, since PQ-learning puts a penalty on each individual, it automatically initiates another important statistical procedure: individual selection. The purpose of individual selection is to select those individuals without treatment effects from the population. Successful individual selection, i.e., correctly identifying individuals without treatment effects, is the key to resolving the non-regularity problem.

Besides improving statistical inference, individual selection is itself an important task in identifying optimal dynamic treatment regimes and in many biomedical and clinical studies. If individuals without treatment effects can be correctly identified, then the corresponding components of the history of these individuals potentially need not be collected to make decisions using the optimal dynamic treatment regime. This could significantly reduce the cost of data collection during implementation of the optimal dynamic treatment regime. While the proposed individual selection procedure shares some similarities with certain commonly used variable selection methods, the approaches are fundamentally different in other ways. These issues will be addressed in greater detail in the paper.

%In contrast to these aforementioned approaches under Q-learning framework, we do not pursue along the direction of regularizing (or shrinking) the ``problematic term'', which involves absolute functional form of some regular estimators. Instead,

%Our contributions build on the previous literature can be casted into several aspects. Firstly, we initiate the concept of individual selection. Secondly, we proposed valid statistical inference procedure for two-stage set-up. The proposed procedure also holds for multi-stage data set-up as will be discussed in the paper. %Our second contribution is that we initiate the concept of individual selection,

The remainder of the paper is organized as follows. In Section 2, we provide a review of statistical problems in reinforcement learning. The proposed penalize Q-learning and individual selection procedure are presented in Section 3, where the implementation and the statistical properties are discussed in detail. Some empirical results are presented in Section 4. We apply the proposed approach to the Sequenced Treatment Alternatives to Relieve Depression (STAR*D) clinical trial in Section 5. A summary of our findings with a discussion is given in Section 6. Proofs are deferred to the Appendix.

\section{Statistical Problems in Reinforcement Learning}
\subsection{Reinforcement Learning}
The basic reinforcement learning procedure involves
\begin{itemize}
\item[i] trying and recording a sequence of actions,
\item[ii] statistically estimating the relationship between the actions and consequences and
\item[iii] choosing the action that results in the most desirable consequence based on statistical decisions.
\end{itemize}
A detailed introduction of reinforcement learning can be found in \cite{Sutton98}. In a reinforcement learning based clinical trial design, we choose a sequence of actions applied to the patient and the environment responds to those actions and provides feedback. Here, ``environment'' refers to the system consisting of the human body and related additional sources of measurements. Specifically, we use random variable $S$ to denote the set of environmental states and $A$ to denote the set of possible actions. For example, states can represent individual patient prognostic factors and actions can represent different treatment agents or dose levels. Their time-dependent versions are denoted $\bS_t = \{S_0, S_1, \ldots S_t\}$ and $\bA_t = \{A_0, A_1, \ldots, A_t\}$, respectively. We use the corresponding lower case to denote a realization of these random variables and random vectors.  The time points correspond to clinical decision points in the course of patient treatment.
After each time step $t$, as a consequence of a patient's treatment, the patient receives a numerical reward, denoted with a random variable $R_t$, which can be represented as a function $R(\cdot)$ of the current state $\bS_t$, current action $\bA_t$ and next state $S_{t+1}$, that is, $R_t = R(\bS_t, \bA_t, S_{t+1})$. We also denote a realization of $R_t$ as
$r_t = R(\bs_t, \ba_t, s_{t+1})$.

Within a reinforcement learning framework, an exploration policy $\pi$
can  be represented as $\pi_t(\bs_t, \ba_{t-1})=a_t$, a mapping from
state $\bs_t$ and action $\ba_{t-1}$ to the set of possible actions. In the clinical setting, a policy is a treatment regimen or a rule. Since our goal in the clinical setting focuses on discovering the treatment that yields a maximized long term reward for the patient, i.e., an optimal personalized treatment, thus seeking the optimal policy that maximizes the expectations of the total rewards over the time trajectories is a major goal of clinical research. Accordingly, we define a value function as a function of states and actions:
\beqn
V_t(\bs_t, \ba_{t-1}) = E_{\pi}\Bigl[\sum_{k=0}^U \gamma^{k}r_{t+k}\Bigl|\bS_t=\bs_t, \bA_{t-1}=\ba_{t-1} \Bigr],\Bigr.
\eeqn
where the discount rate $\gamma \in [0,1]$ can be interpreted as a control to balance a patient's immediate reward and future rewards. The value function measures the success of the treatment policy $\pi$. Letting $\Pi$ denote the set of all policy candidates, the optimal value function can be defined as
$V_t^{\star}(\bs_t, \ba_{t-1}) = \max_{\pi \in \Pi} V_t(\bs_t, \ba_{t-1}).$

The value functions used in reinforcement learning typically satisfy the recursive Bellman equation (Bellman, 1957), which forces the optimal policy $\pi_t^{\star}$ to satisfy
\beqn
\pi_t^{\star}(\bs_t, \ba_{t-1}) \in \argmax_{a_t} E\Bigl[ r_t + \gamma V_{t+1}^{\star}(\bS_{t+1}, \bA_t) \Bigl|\bS_t=\bs_t, \bA_{t-1}=\ba_{t-1} \Bigr].\Bigr.
\eeqn
Due to computational challenges, it is usually not possible to directly compute an optimal policy by directly solving the Bellman equation. As an alternative method which requires less memory and less computation, temporal-difference learning can also be used to obtain optimal policies \citep{Sutton88,KMA96}. In the next section, we will introduce a very important off-policy temporal-difference learning method, Q-learning, which is a  popular approach to estimate dynamic treatment regimes. Q-learning is the estimating approach for which the statistical inference procedures in \cite{Cha09} were proposed.

\subsection{Q-Learning Procedure}
The motivation of Q-learning is that once the Q-functions have been estimated, it is only necessary to know the state to determine the best action. From a statistical perspective, the optimal time-dependent Q-function is
\beqn
Q_t^{\star}(\bs_t, \ba_t) = E\Bigl[ r_t +\gamma V_{t+1}^{\star}(\bS_{t+1}, \bA_t)\Bigl|\bS_t=\bs_t, \bA_{t}=\ba_{t} \Bigr].\Bigr.
\eeqn
Since by definition, $V_{t+1}^{\star}(\bs_{t+1}, \ba_{t})= \max_{a_{t+1}}Q_{t+1}^{\star}(\bs_{t+1}, \ba_{t}, a_{t+1}),$
and hence \\
$\pi_{t+1}^{\star}(\bs_{t+1}, \ba_{t})=\argmax_{a_{t+1}}Q_{t+1}^{\star}(\bs_{t+1}, \ba_{t}, a_{t+1}),$
one-step Q-learning thus has the simple recursive form:
\beq\label{eq-Q}
Q_{t}(\bs_t, \ba_t) = E\Bigl[R_t + \gamma \max_{a_{t+1}}Q_{t+1}(\bS_{t+1}, \ba_{t+1})\Bigl|\bS_t = \bs_t,
\bA_t = \ba_t \Bigr]. \Bigr.
\eeq
%It can be shown that under certain conditions, $Q_t$ converges to $Q_t^{\star}$ with probability one (Watkins, 1992).
According to the recursive form of Q-learning in (\ref{eq-Q}), we must estimate $Q_t$ backwards through time $t=U,U-1,\ldots,0$. To estimate each $Q_t$, we parameterize $Q_t(\bs_t, \ba_t; \btheta_t)$ as a function of the parameter $\btheta_t$. After finishing estimation through this backward recursive process and obtaining the sequence estimators $\{\widehat Q_t \}_{t=0}^U$, we can estimate the optimal treatment regimes $\widehat \pi_t = \argmax_{a_t}\widehat Q_t(\bs_t, \ba_t; \btheta_t),$ for $t=1,\ldots,U.$

We will use a simple, two-stage example (also used in \cite{Cha09})
to illustrate the proposed approaches. Let the Q-function for time
$t=1,2$ be modeled as
\beq\label{eq-Q-lse}
Q_t(\bS_t, A_t;
\bbeta_t, \bpsi_t) = \bbeta_t^T\bS_{t1} + (\bpsi_t^T\bS_{t2})A_t,
\eeq
where $\bS_t$ is the full state information at time $t$ and
$\bS_{t1}$ and $\bS_{t2}$ are subsets of $\bS_t$ selected for the
model. The action $A_t$ takes value $1$ or $-1$.
The parameters of the Q-function are $\btheta_t=(\bbeta_t^T,\bpsi_t^T)^T$,
where $\bbeta_t$ reflects the main effect of current
state on outcome, while $\bpsi_t$ reflects the interaction effect
between current state and treatment choice. Let $Y_{t}$ denote the
optimal total potential reward at time $t$. In this work, we will
assume that $Y_{2}=R_{2}$, the discount rate $\gamma=1$  and
$Y_{1} = R_{1} + \max_{a \in \{ -1,1\}} Q_2(\bS_{2}, a;
\bbeta_{20}, \bpsi_{20})$, where $\bbeta_{20}$ and $\bpsi_{20}$
are the true unknown values. The observed data consist of
$(\bS_{ti}, A_{ti}, R_{ti})$ for patients $i=1,\ldots,n$ and
$t=1,2$, from a sample of $n$ patient trajectories.

The two-stage empirical version of the Q-learning procedure can now be
summarized as follows:
\begin{enumerate}
\item[Step 1.] Start with a regular and non-shrinkage estimator, based on least squares, for the second stage:
\begin{align*}
\hbtheta_2 = (\widehat \bbeta_2^T, \widehat \bpsi_2^T)^T  = \argmin_{\bbeta_2, \bpsi_2} \bbP_n (Y_2 - Q_2(\bS_2, A_2; \bbeta_2, \bpsi_2))^2= \Bigl [\bZ_{2}^T\bZ_{2} \Bigr]^{-1} \bZ_{2}^T\bY_{2},
%&= \Bigl [n^{-1}\sum_{i=1}^n \bZ_{2i}\bZ_{2i}^T \Bigr]^{-1} n^{-1}\sum_{i=1}^n \bZ_{2i}Y_{2i},
\end{align*}
where $\bZ_{2}$ is the stage-2 design matrix and $\bY_2=(Y_{21},...,Y_{2n})^T$ and $\bbP_n f(x) = 1/n \sum_{i=1}^n f(x_i)$ is the empirical measure.
\item[Step 2.] Estimate the first-stage individual pseudo-outcome by $\widehat \bY_1^{HM}=(\widehat Y_{11}^{HM}, \ldots, \widehat Y_{1n}^{HM})^T$, where
\beq\label{eq-hm}
\widehat Y_{1i}^{HM} &= R_{1i} + \max_{a}Q_2(\bS_{2i}, a; \widehat \bbeta_2, \widehat \bpsi_2) =R_{1i} + \widehat \bbeta_2^T \bS_{21i} + |\widehat \bpsi_2^T \bS_{22i}|.
\eeq

\item[Step 3.] Estimate the first-stage parameters by least square estimation:
\begin{align*}
\hbtheta_1^{HM} = \argmin_{\bbeta_1, \bpsi_1} \bbP_n (\widehat Y_1^{HM} - Q_1(\bS_1, A_1; \bbeta_1, \bpsi_1))^2 =\Bigl [\bZ_{1}^T\bZ_{1} \Bigr]^{-1} \bZ_{1}^T\widehat\bY_{1}^{HM},
\end{align*}
where $\bZ_{1}$ is the stage-1 design matrix.
The corresponding estimator of $\bpsi_1$, denoted by $\widehat \bpsi_1^{HM}$, is referred to as the hard-max estimator in \cite{Cha09}, because of the maximizing operation used in the definition.
\end{enumerate}

\subsection{Non-regularity Problem in Statistical Inference}
When the Q-function is taking the linear model form~(\ref{eq-Q-lse}), the optimal dynamic treatment regime for patient $\{i, i=1,\ldots,n\}$ is given by
\beqn
d_i(\bS_{ji})=\argmax_{a_i}(\bpsi_j^T\bS_{j2i})a_i = \sgn(\bpsi_j^T\bS_{j2i}),~\mbox{for}~j=1,2,
\eeqn
where $\sgn(x)=1$ if $x>0$ and $-1$ otherwise. The parameters $\bpsi_2$ are of particular interests for estimation and inference of the optimal dynamic treatment regime, as confidence intervals for $\bpsi_j$ can lead to confidence intervals for $d_i$.

During the Q-learning procedure, when there is a positive probability that $\bpsi_{20}^T\bS_{22}=0$, the first-stage hard-max pseudo-outcome $\widehat \bY_1^{HM}$ is a non-smooth function of $\hbpsi_2$. As a linear function of $\widehat \bY_{1}^{HM}$, the hard-max estimator $\hbpsi_1^{HM}$ is also a non-smooth function of $\widehat \bpsi_2$. Consequently, the asymptotic distribution of $\sqrt{n}(\hbpsi_1^{HM}-\bpsi_{10})$ is neither normal nor any well-tabulated distributions if $P(\bpsi_{20}^T\bS_{22}=0)>0$.  In this non-standard case, standard inference methods such as Wald-type confidence intervals are no longer valid.

\subsection{Review of Existing Approaches}
To overcome the difficulty of inference of $\bpsi_1$ due to non-regularity in Q-learning, several methods have been proposed and we will briefly review these methods in the two-stage set-up. They are referred to as the hard-threshold estimator (also called {\it Zeroing Instead of Plugging In} (ZIPI) estimator in Moodie and Richardson (2010) and the soft-threshold estimator in \cite{Cha09}. Since all these methods are also nested in the Q-learning procedure, we update the two-stage version of Q-learning as follows:
\begin{enumerate}
\item[Step 2'.] Estimate the first-stage individual pseudo-outcome by shrinking the second-stage regular estimators via hard-thresholding or soft-thresholding. Specifically, the hard-threshold pseudo-outcome is denoted $\widehat \bY^{HT}_1=(\widehat Y^{HT}_{11},\ldots,\widehat Y^{HT}_{1n})^T$, with
\beq\label{eq-ht}
\widehat Y^{HT}_{1i}  = R_{1i} + \hbbeta_2^T \bS_{21i} + |\hbpsi_2^T \bS_{22i}|\cdot 1\Bigl\{
\frac{\sqrt{n}|\hbpsi_2^T\bS_{22i}|}{\sqrt{\bS_{22i}^T\widehat\Sigma_2\bS_{22i}}}>z_{\alpha/2}\Bigr\},
\eeq
where $\widehat \Sigma_2$ is the estimated covariance matrix of $\hbpsi_2$.

The soft-threshold pseudo-outcome is denoted $\widehat \bY_1^{ST} = (\widehat Y_{11}^{ST}, \ldots, \widehat Y_{1n}^{ST} )^T$, with
\beq\label{eq-st}
\widehat Y_{1i}^{ST} = Y_{1i} + \widehat \bbeta_2^T \bS_{21i} + |\widehat \bpsi_2^T \bS_{22i}|\Bigl(1-\frac{\lambda_i}{|\widehat \bpsi_2^T \bS_{22i}|} \Bigr)^+,~i=1,\ldots,n,
\eeq
where $x^+ = xI\{x>0 \}$ is the positive part of a function and $\lambda_i$ is a tuning parameter.

\item[Step 3'.] Estimate the first-stage parameters by least squares estimation:
\begin{align*}
\hbtheta_1^T=  (\widehat \bbeta_1^T, \widehat \bpsi_1^T)^T  = \argmin_{\bbeta_1, \bpsi_1} \bbP_n (\widehat Y_1^{\circ} - Q_1(\bS_1, A_1; \bbeta_1, \bpsi_1))^2 =\Bigl [\bZ_{1}^T\bZ_{1} \Bigr]^{-1} \bZ_{1}^T\widehat\bY_{1}^{\circ},
\end{align*}
where $\widehat\bY_1^{\circ}=(\widehat Y_{11}^{\circ},...,\widehat Y_{1n}^{\circ})^T$ is the first-stage pseudo-outcome obtained in Step 2'. It can be either a hard-threshold or soft-threshold pseudo-outcome, as defined in either (\ref{eq-ht}) or (\ref{eq-st}), respectively.
The corresponding estimator of $\bpsi_1$, denoted $\widehat \bpsi_1^{\circ}$, can be
the hard-threshold estimator $\widehat \bpsi_1^{HT}$ (also referred to as {\it Zeroing Instead of Plugging} (ZIPI) estimator in \cite{17Moodie08}) or the soft-threshold estimator $\widehat \bpsi_1^{ST}$, respectively.
\end{enumerate}

%When there is a positive probability such that $\hbpsi_2^T\bS_{22}=0$ for least square estimator $\hbpsi_2$, the first-stage hard-max pseudo-outcome $\widehat \bY_1^{HM}$ is a non-smooth function of $\hbpsi_2$. As a linear function of $\widehat \bY_{1}^{HM}$, the hard-max estimator $\hbpsi_1^{HM}$ is also a non-smooth function of $\widehat \bpsi_2^{HM}$. Consequently, the asymptotic distribution of $\sqrt{n}(\hbpsi_1^{HM}-\bpsi_{10})$ is neither normal nor any well-tabulated distributions if $P(\hbpsi_2^T\bS_{22}=0)>0$.  In this non-standard case, standard inference methods such as Wald-type confidence intervals are not valid any more.

The hard-thresholding and soft-thresholding methods can be viewed as upgraded versions of the hard-max methods in terms of reducing the degree of non-regularity. For example, \cite{Cha09} commented that the third term in (\ref{eq-st}) takes the form of the non-negative garrote estimator \citep{Breiman95}, through which the problematic term $|\hbpsi_2^T \bS_{22}|$ is expected to shrink (or thresholded) towards zero.  %even if $\widehat Y_1^{ST}$ is also a non-smooth function of $\widehat \bpsi_2$.
Even if the degree of non-regularity is somewhat decreased, in general, $\widehat \bpsi_1^{HT}$ and $\widehat \bpsi_1^{ST}$ remain non-regular estimators of $\bpsi_{10}$, and standard inference methods such as Wald-type confidence intervals are still not valid. Numerical studies in \cite{Cha09}  show an apparent  bias for these two methods in certain simulation settings.

Due to the nature of the non-regularity, the asymptotic distributions of these three estimators are not well tabulated, and thus direct inference for $\bpsi_1$ is not feasible. Bootstrap methods seem to be the only remedy.  \cite{Cha09} applied several bootstrap methods to construct confidence interval for $\bpsi_1$. Unfortunately, no theoretical support was provided for any of these methods. Moreover, none of these bootstrap confidence intervals for hard-thresholding and soft-thresholding methods perform uniformly well in all of the simulation scenarios considered.

In summary, the first-stage pseudo-outcome for these three existing methods can be viewed as
shrinkage functionals of certain standard estimators (such as least square estimator in this two-stage set-up). Although the idea of simultaneously approximating the hard-max estimator (or
more precisely, the absolute value function) and reducing the
degree of non-regularity sounds reasonable, it may not be an
appropriate approach to use shrinkage formulations of existing
regular estimators directly as the first-stage pseudo-outcome. The
reason is that even if these estimators form shrinkage estimators under certain conditions (e.g., least squares regression with the only covariates $\bS_{21}$ and under an orthonormal design), in general, they are not optimizers of reasonable objective functions. Consequently, even if these estimators can successively achieve the goal of shrinkage, the following two drawbacks remain that negate their ability to be used for statistical inference for optimal
dynamic treatment regimes.

First, the bias of these ``shrinking'' first-stage
pseudo-outcomes can be large in finite samples, leading to
further bias in the first stage estimation of $\bpsi_1$. This
point has been demonstrated in the empirical studies of
\cite{Cha09}. Second, and more importantly, these shrinking functional estimators do not
appear to possess the oracle property. Here we refer to the oracle property
as: with probability tending to one, the set $\mathcal{M}_{\star} = \{i:
|\bpsi_{20}^T \bS_{22i}|>0 \}$ can be correctly identified and the
resulting estimator performs as well as the oracle estimator,
which knows in advance the right set $\mathcal{M}_{\star}$. These two concerns
probably are the direct reasons why these three existing
estimation methods, with their corresponding bootstrapping
confidence intervals, do not perform well (with bias and invalid
coverage length, for example) in some empirical studies. This suggests that
it may not be appropriate to attempt to fix the non-regularity by regularizing existing estimators directly.

% talk about first-stage estimator, second-stage estimator.
\section{Inference Based on Penalized Q-Learning}

In this section, we propose an innovative method, penalized Q-learning (PQ-learning), for statistical inference in reinforcement learning problems. Our method is based
on individual selection via a penalized likelihood; so it will
automatically determine those individuals whose value functions
are not affected by treatment assignments, i.e, those individuals for whom
$\bpsi_2^T\bS_{22}$ given in Section 2.2 is $0$. As a result,
this oracle property ensures that subsequent inference in the
Q-learning framework will be the same as if we knew which
individuals had no treatment effect.
Accordingly, the resulting inference will no longer suffer from the
non-regularity problem discussed above. To describe our
method, we still focus on the two-stage setting as given in
Section 2.2 and use the same notation. The generalization
to the multiple-stage setting is similar and we will it discuss briefly
at the end of the section.

\subsection{Estimation Procedure}
As a backward recursive reinforcement learning procedure, our method follows essentially the same three steps as the usual Q-learning method. The only difference is that our approach replaces Step 1 of the standard Q-learning procedure with the following:

\begin{enumerate}
\item[Step 1'.] Instead of minimizing the summed squared differences between
$Y_2$ and $Q_2(\bS_2, A_2;\bbeta_2, \bpsi_2)$, we minimize the
following penalized objective function: \beq\label{eq-penalty}
W_2(\btheta_2) = \sum_{i=1}^n(Y_{2i} - Q_2(\bS_{2i}, A_{2i};
\bbeta_2, \bpsi_2))^2 +
\sum_{i=1}^np_{\lambda_n}(|\bpsi_2^T\bS_{22i}|), \eeq where
$p_{\lambda_n}(\cdot)$ is a pre-specified penalty function and
$\lambda_n$ is a tuning parameter.
\end{enumerate}

Because of this penalized estimation, we call our approach
``penalized Q-learning'' and abbreviate it as
PQ-learning. Since the penalty is put on each individual, we also call Step 1' ``individual selection.''

Using penalty functions in PQ-learning, i.e., individual selection, enjoys similar ``shrinkage''
advantages as penalized methods described in the recent variable selection literature.
To name a few, the bridge regression in \cite{FF93}, the LASSO in \cite{Tibs96}, the
SCAD and other folded-concave penalties in \cite{FanLi01}, the
Dantzig selector in \cite{CT07}, the adaptive LASSO in \cite{Zou06} and one-step estimator in \cite{ZouLi08}. In variable selection problems where the selection of interest consists of the important variables (mostly covariates) with nonzero coefficients, using appropriate penalties can shrink the small estimated coefficients to zero to enable the desired selection. In the individual selection done in the first step of the proposed PQ-learning approach, penalized estimation allows us to simultaneously estimate the second-stage parameters $\btheta_2$ and select individuals whose value functions are not affected by treatments, i.e., those individuals whose true values of $\bpsi_2^T\bS_{22}$ are zero.

The above fact is extremely useful in making correct inference in
the subsequent steps of the PQ-learning procedure. To understand
why, we recall that the non-regularity problem in the usual
Q-learning procedure is mainly caused by difficulties in obtaining the correct
asymptotic distribution of
$$\sqrt n (|\hbpsi_2^T\bS_{22}|-|\bpsi_{20}^T\bS_{22}|),$$
where $\bpsi_{20}$ is the true value of $\bpsi_2$. Via our PQ-learning
method, we can identify individuals whose
$\widehat\bpsi_2\bS_{22}=\bpsi_{20}^T\bS_{22}$ takes value zero; moreover, we
know that for these individuals $\widehat\bpsi_2^T\bS_{22}$ has the same sign as
$\bpsi_{20}^T\bS_{22}$. In this case, the above expression is equivalent to
$$\sqrt n (\widehat\bpsi_2 -\bpsi_{20})^T\bS_{22}
\textrm{sign}(\bpsi_0^T\bS_{22}).$$
Hence, correct inference can be obtained following standard arguments. More rigorous details will
be given in our subsequent asymptotic theorems and proofs.

The choice of the penalty function $p_{\lambda_n}(\cdot)$ can be taken to be the
same as used many popular variable selection methods. Specifically, we
require $p_{\lambda_n}(\cdot)$ to possess the following properties:
\begin{itemize}
\item[A1.] For non-zero fixed $\theta$, $\lim_{n\rightarrow \infty} n^{1/2}p_{\lambda_n}(|\theta|) =0$,
$\lim_{n \rightarrow \infty} n^{1/2} p'_{\lambda_n}(|\theta|)=0$, \\
    and $\{p_{\lambda_n}{''}(|\theta|) \} \rightarrow 0$.
\item[A2.] For any $M>0$, $\lim_{n \rightarrow \infty} \inf_{|\theta|\le M n^{-1/2}} p_{\lambda_n}(|\theta|) \rightarrow \infty$.
\end{itemize}
Among many penalty functions satisfying A1 and A2, some common
choices include the SCAD penalty \citep{FanLi01}
%where $$p'_{ \lambda_n}(\theta) = \lambda_n\Bigl\{I(\theta \le \lambda_n)
%+(a\lambda_n-\theta)^+/\{(a-1)\lambda_n\} I(\theta>\lambda_n) \Bigr\},$$
and the adaptive lasso penalty \citep{Zou06}, where $p_{\lambda_n}(\theta) =
{\lambda_n\theta}/{|\theta^{(0)}|^\alpha}$ with $\alpha>0$
and $\theta^{(0)}$ being a root-$n$ consistent estimator of
$\theta$. To achieve both sparsity and oracle properties, the tuning parameter $\lambda_n$ in these examples should be taken correspondingly. For illustration, Figure 1 plots these penalty functions. The adaptive lasso method will be implemented in this paper, where $\lambda_n$ can be taking as scalars satisfying $\sqrt{n}\lambda_n \rightarrow 0$ and
$n\lambda_n \rightarrow \infty$.

\begin{figure}[htbp]
%[htbp]
\centering
%\hspace{0.033\textwidth}
\rotatebox{270}{
\makebox{\includegraphics[width=3in, height=6.5in]{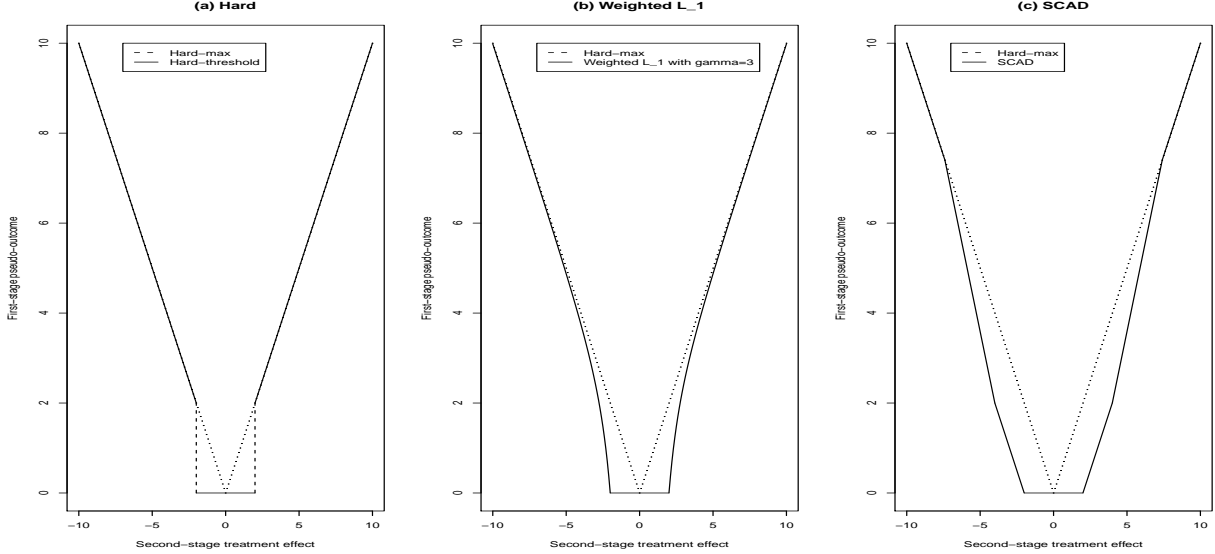}}}
   \caption{Plot of thresholding functions with $\lambda$=2 for (a) the Hard Threshold;
   (b)adaptive lasso with $\alpha=3$; (c) SCAD, with the hard-max function as a reference in each panel.}
\label{f2}
% \hspace{0.033\textwidth}
%\hfill
\end{figure}

\subsection{Implementation}
The minimization in Step 1' of the PQ-learning procedure has some
unique features which distinguish it from the optimization done
in the variable selection literature. First, the component to be
shrunk, $\bpsi_2^T\bS_{22i}$, is subject-specific; second, this
component is a hyperplane in the parameter space, i.e, a linear
combination of the parameters.

To deal with these issues, in this section, we propose an
algorithm for the minimizing problem of (\ref{eq-penalty}) based on
local quadratic approximation (LQA). Following Fan and Li (2001),
we first calculate an initial estimator $\widehat\bpsi_{2(0)}$. We then
obtain the following LQA to the penalty terms in (\ref{eq-penalty}):
\[
p_{\lambda_n}(|\bpsi_2^T\bS_{22i}|) \approx
p_{\lambda_n}(|\widehat\bpsi_{2(0)}^T\bS_{22i}|) +
\frac{1}{2}\frac{p'_{\lambda_n}(|\widehat\bpsi_{2(0)}^T\bS_{22i}|)}{|\widehat\bpsi_{2(0)}^T\bS_{22i}|}((\bpsi_2^T\bS_{22i})^2-
(\widehat\bpsi_{2(0)}^T\bS_{22i})^2)
\]
for $\bpsi_2$ close to $\widehat\bpsi_{2(0)}$. Thus, (\ref{eq-penalty})
can be locally approximated up to a constant by \beq\label{eq-lqa}
 \sum_{i=1}^n(Y_{2i} - Q_2(\bS_{2i}, A_{2i};
\bbeta_2, \bpsi_2))^2 +
\frac{1}{2}\sum_{i=1}^n \frac{p'_{\lambda_n}(|\widehat\bpsi_{2(0)}^T\bS_{22i}|)}{|\widehat\bpsi_{2(0)}^T\bS_{22i}|}(\bpsi_2^T\bS_{22i})^2.
%||\bY_2-
%Q(\bS_2, A_2; \bpsi_2, \bbeta_2)||^2 +
%\bpsi_2^T\bS_{21}\bD\bS_{21}^T\bpsi_2,
\eeq
The updated estimators for $\bpsi_2$ and $\bbeta_2$ can be
obtained by minimizing the above approximation. Under the special
case where $Q(\cdot)$ is given by (2), this minimization problem
has a closed form solution:
\[ \widehat\bpsi_2 =\big[\bX_{22}^T(\bI-\bX_{21}(\bX_{21}^T\bX_{21})^{-1}\bX_{21}^T
+ \bD)\bX_{22}\big]^{-1}\bX_{22}^T(\bI-\bX_{21}(\bX_{21}^T\bX_{21})^{-1}\bX_{21}^T)\bY_2,\]
\[\widehat\bbeta_2=(\bX_{21}^T\bX_{21})^{-1}\bX_{21}^T(\bY_2-\bX_{22}\widehat\bpsi_2),\]
where $\bX_{22}$ is a matrix with $i$-th row equal to
$\bS_{22i}^TA_{2i}$, $\bX_{21}$ is a matrix with $i$-th row equal to
$\bS_{21i}^T$, $\bI$ is the $n\times n$ identity matrix and $\bD$ is an
$n\times n$ diagonal matrix with
$D_{ii}=\frac{1}{2}p_{\lambda_n}'(|\widehat\bpsi_{2(0)}^T\bS_{22i}|)/|\widehat\bpsi_{2(0)}^T\bS_{22i}|$.

The above minimization procedure can be continued for more than
one step or until convergence. However, as discussed in Fan and Li
(2001), either the one-step or $k$-step estimator will be as efficient
as the fully iterative method as long as the initial estimators
are good enough. A well known limitation of the LQA algorithm is that
although it can shrink $|\widehat\bpsi_2^T\bS_{22i}|$ to a very small
value if the true value is zero, it cannot set it exactly to zero. Therefore, in practice, we will set $|\widehat\bpsi_2^T\bS_{22i}|=0$ once the value is below a
pre-specified tolerance threshold.

The choice of LQA is mainly for convenience in solving the penalized
least squares estimation in (\ref{eq-penalty}). If least absolute
deviation estimation or some other quantile regressions is used in place
of least squares, then the local linear approximation
of the penalty function described in \cite{ZouLi08} can be used instead of LQA, and
the resulted minimization problem can be solved by linear programming. The linear
programming approach can shrink small values of $|\widehat\bpsi_2^T\bS_{22i}|$ exactly to zero
and therefore avoids the additional thresholding done in LQA. Alternatively, the
Dantzig selector \citep{CT07} can be used with penalized least squares estimation
in (\ref{eq-penalty}), but the asymptotic properties for this setting are beyond the scope of this paper, although they are currently being investigated by the authors.

\subsection{Asymptotic Results}

In this section, we establish the asymptotic properties for the
parameter estimators in our PQ-learning method, assuming that the
support of $\bS_{22}$ contains a finite number of vectors, say,
$\bT_1,...,\bT_K$. Moreover, we assume $\bpsi_{20}^T\bT_k\neq 0$
for $k\le K_1$ and $\bpsi_{20}^T\bT_k= 0$ for $k>K_1$. Let
$$n_k=\#|\{i:~\bS_{22i}=\bT_k,~i=1,\ldots,n\}|,$$
where for a set $A$, $\#|A|$ is defined as its cardinality.

Additionally, we assume that the penalty function $p_{\lambda_n}(x)$
satisfies A1 and A2 and that the following conditions hold:
\begin{enumerate}
\item[B1.] The true value for $\btheta_2$, denoted by
$\btheta_{20}=(\bpsi_{20}^T, \bbeta_{20}^T)^T$, minimizes
$$\lim_n \bbP_n\left[Y_2-Q_2(\bS_2, A_2;\bbeta_2, \bpsi_2)\right]^2;$$
while, the true value for $\btheta_1$, denoted by
$\btheta_{10}=(\bpsi_{10}^T, \bbeta_{10}^T)^T$, minimizes
$$\lim_n \bbP_n\left[R_1+\max_{a}Q_2(\bS_2, a; \bbeta_{20},
\bpsi_{20})-Q_1(\bS_1, A; \bbeta_1, \bpsi_{1})\right]^2.$$ In both
expressions and in the following, we always assume that the limits
exist.

\item[B2.] For $k=1,2$, with probability one, $Q_k(\bS_k, A_k;
\btheta_k)$ is twice-continuously differentiable with respect to
$\btheta_k$ in a neighborhood of $\btheta_{k0}$ and moreover, the
eigenvalues of the Hessian matrix, $I_{k0}\equiv \lim_n \bbP_n
[\nabla_{\btheta_k\btheta_k}^2 (Y_k-Q_k(\bS_k, A_k;
\btheta_k))^2]$, are positive and bounded away from zero at
$\btheta_k=\btheta_{k0}$.

\item[B3.] With probability one, $n_k/n=p_k+O_p(n^{-1/2})$ for some
constant $p_k$ in $[0,1]$.

\end{enumerate}
Condition B1 basically says that $\btheta_{10}$ and
$\btheta_{20}$ are the target values in the dynamic treatment
regimes, which we consider to be the true values. Condition B2 can be
verified via the design matrix in the two-stage setting.
Specifically, if $Q_t$ takes the form of (2), this condition is
equivalent to non-singularity of the design matrix $[A_t,
\bS_{t}A_t]$.

Under these conditions, our first theorem shows that in Step 1' of
the PQ-learning procedure, there exists a consistent estimator for
$\btheta_2$:
\begin{theorem}\label{the1}
Under conditions A1-A2 and B1-B3, there exists a local minimizer
$\hbtheta_2$ of $W_2(\btheta_2)$ such that $\|\hbtheta_2 -
\btheta_{20} \| = O_P(n^{-1/2} + a_n)$, where
$a_n=\max_{k=1}^{K_1} \left\{p'_{\lambda_n}(|\bpsi_{20}^T\bT_k|)\right\}.$
\end{theorem}

According to the properties of $p_{\lambda_n}(\cdot)$, we
immediately conclude that $\hbtheta_2$ is $\sqrt n$-consistent.
From Theorem 1, we further obtain the following result, which
verifies the oracle property of the penalized method:
\begin{theorem}\label{the2}
Recall the set
$${\cal M}_{\star}^c=\left\{i: \bpsi_{20}^T\bS_{22i}=0\right\}.$$
Then under conditions A1-A2 and B1-B3, \beqn \lim_{n \rightarrow
\infty}P(\hbpsi_2^T\bS_{22i}=0,~ \mbox{for any}~i \in {\cal M}_{\star}^c) = 1.
\eeqn
\end{theorem}

The set ${\cal M}_{\star}^c$ consists of those individuals whose true value
functions at the first stage have no effect from treatment. Thus Theorem 2 states that with probability tending to one, we can identify these individuals in ${\cal M}_{\star}^c$ using empirical
observations. As discussed before, this result will be very useful
in addressing the non-regularity problem for subsequent inference. Additionally,
we will also need the asymptotic distribution of $\hbtheta_2$
in order to make inference. This is provided in the following theorem:
\begin{theorem}\label{the3}
Under conditions A1-A2 and B1-B3,
\beq \sqrt{n}(I_{20} +
\Sigma)\{\hbtheta_{2} - \btheta_{20} + (I_{20} + \Sigma)^{-1}\bb \}
\rightarrow N\{0, I_{20} \}, \eeq where
$$\bb= \Bigl(\bzero_p^T, \sum_{k=1}^{K_1} p_k p_{\lambda_n}'(|\bpsi_{20}^T\bT_k|)\sgn(\bpsi_{20}^T\bT_k)\bT_K\Bigr)^T, $$
$$\mbox{and}~\Sigma=  \mbox{diag}\{\bzero_{p\times p}, \sum_{k=1}^{K_1} p_k p_{\lambda_n}''(|\bpsi_{20}^T\bT_k|)\bT_k\bT_k^T\}. $$
\end{theorem}

Using the results from Theorems 1--3, we are able to
establish asymptotic normality of the first stage estimator $\hbtheta_1$:
\begin{theorem}\label{the4}
Under conditions A1-A2 and B1-B3, let $\bar\bS_2 \equiv (\bS_{21}^T, \bS_{22}^T\sgn(\bpsi_{20}^T \bS_{22}))^T .$ Then
\begin{eqnarray}\label{eq-asy}
\sqrt{n}
(\hbtheta_1 - \btheta_{10})
\rightarrow
I_{10}^{-1}
\bigl\{
G_1 +
  \lim_n \bbP_n \bZ_1\bar\bS_2^T  G_2\bigr\},
\end{eqnarray}
where
$G_1 \sim N\Bigl[0, \lim_n \cov\left\{\nabla_{\btheta_1}Q_1(\bS_{1},
A_1;\btheta_{10})(Y_1-Q_1(\bS_{1}, A_1;
\btheta_{10}))\right\} \Bigr]$,\\
$G_2 \sim N\Bigl[0, (I_{20} + \Sigma)^{-1}I_{20}(I_{20} + \Sigma)^{-1} \Bigr]$ and $\cov$ represent the sample variance. %$G_1$ and $G_2$ are independent.

%$I_{22.1}^{-1}$ is the lower submatrix of $I_{22}^{-1}$ consisting of the bottom $p$ rows and $I_{22.0}^{-1}$ is the upper submatrix consisting of the remaining rows of $I_{22}^{-1}$.
\end{theorem}

\subsection{Variance Estimation}
The standard errors for the estimated parameters can be obtained directly since we are estimating parameters and selecting individuals simultaneously. A sandwich type plug-in estimator can be used as the variance estimator for $\hbtheta_{2}$:
\beqn
\widehat {\cov}(\hbtheta_2) = (\widehat I_{20} + \widehat \Sigma)^{-1}\widehat I_{20}(\widehat I_{20} + \widehat \Sigma)^{-1},
\eeqn
where $\widehat I_{20}\equiv \bbP_n
[\nabla_{\btheta_2\btheta_2}^2 (Y_2-Q_2(\bS_2, A_2;
\btheta_2))^2]$ is the empirical Hessian matrix and
$\widehat \Sigma =
 \mbox{diag}\{\bzero_{p\times p}, \bbP_n p_{\lambda_n}''(|\hbpsi_{2}^T\bS_{22}|)\bS_{22}\bS_{22}^T\}.$
As $\widehat \Sigma$ is often negligible, we use
\beq\label{eq-var-1}
\widehat {\cov}(\hbtheta_2) = \widehat I_{20}^{-1}
\eeq
instead, and this performs well in practice. The estimated variance for $\hbtheta_1$ is then
$\widehat {\cov}(\hbtheta_1) = $
\beq\label{eq-var-2}
\widehat I_{10}^{-1}\Bigl[ \cov\left\{\nabla_{\btheta_1}Q_1(\bS_{1},
A_1;\hbtheta_{1})(Y_1-Q_1(\bS_{1}, A_1;
\hbtheta_{1}))\right\} + \bbP_n \bZ_1\bar {\bS}_2^T \widehat {\cov}(\hbtheta_2) \bar {\bS}_2 \bZ_1^T  \Bigr]\widehat I_{10}^{-1},
\eeq
where $\widehat I_{10}\equiv \bbP_n
[\nabla_{\btheta_1\btheta_1}^2 (Y_1-Q_1(\bS_1, A_1;
\btheta_1))^2]$ is the empirical Hessian matrix.
These variance estimators will be shown in simulations presented later to have good accuracy for moderate sample sizes. This success of direct inference for the estimated parameters makes statistical inference for optimal dynamic treatment regime possible in the multi-stage setting.
%$I_{k0}\equiv \lim_n \bbP_n [\nabla_{\btheta_k\btheta_k}^2 (Y_k-Q_k(\bS_k, A_k; \btheta_k))^2]$

\subsection{Generalization to the Multi-stage Setting}
In this section, we will extend the inference procedure from the two-stage to the more general multi-stage setting.
The PQ-learning procedure for finding the optimal dynamic treatment regime for a general $U$-stage setting can be summarized in $2U+1$ steps as follows:

\begin{enumerate}
\item[Step 1.] Start from the $U$th stage by minimizing the penalized Q-function at the $U$th-stage:
\beqn
W_U(\btheta_U) = \sum_{i=1}^n(Y_{Ui} - Q_U(\bS_{Ui}, A_{Ui};
\bbeta_U, \bpsi_U))^2 +
\sum_{i=1}^np_{\lambda_n}(|\bpsi_U^T\bS_{U2i}|), \eeqn where
$p_{\lambda_n}(\cdot)$ is a pre-specified penalty function and
$\lambda_n$ is a tuning parameter.

\item[Step 2.] Estimate the $(U-1)$th-stage individual pseudo-outcome by $\widehat \bY_{U-1}=(\widehat Y_{U-1,1}, \ldots, \widehat Y_{U-1, n})^T$, where
\beqn
\widehat Y_{U-1,i} = R_{U-1,i} + \max_{a}Q_{U}(\bS_{U,i}, a; \hbtheta_{U-1}) =R_{U-1,i} + \widehat \bbeta_{U}^T \bS_{U1i} + |\widehat \bpsi_{U}^T \bS_{U2i}|.
\eeqn

\item[Step 3.] Minimize the penalized Q-function in the $(U-1)$th-stage with the individual pseudo-outcome obtained from Step 2:
\beqn
W_{U-1}(\btheta_{U-1}) = \sum_{i=1}^n(\widehat Y_{U-1,i} - Q_{U-1}(\bS_{U-1,i}, A_{U-1,i};
\btheta_{U-1}))^2 +
\sum_{i=1}^np_{\lambda_n}(|\bpsi_{U-1}^T\bS_{U-1,2i}|).
 \eeqn

\item[Step 4.] Estimate the $(U-2)$th-stage individual pseudo-outcome by $\widehat \bY_{U-2}=(\widehat Y_{U-2,1}, \ldots, \widehat Y_{U-2, n})^T$, where
\beqn
\widehat Y_{U-2,i} = R_{U-2,i} + \max_{a}Q_{U-1}(\bS_{U-1,i}, a; \hbtheta_{U-2}) =R_{U-2,i} + \widehat \bbeta_{U-1}^T \bS_{U-1,1i} + |\widehat \bpsi_{U-1}^T \bS_{U-1,2i}|.
\eeqn
\item[$\cdots$]
\item[Step 2U+1.] Estimate the first-stage parameters by least squares estimation:
\begin{align*}
\hbtheta_1 = \argmin_{\bbeta_1, \bpsi_1} \bbP_n (\widehat Y_1 - Q_1(\bS_1, A_1; \bbeta_1, \bpsi_1))^2 =\Bigl [\bZ_{1}^T\bZ_{1} \Bigr]^{-1} \bZ_{1}^T\widehat\bY_{1}.
\end{align*}
\end{enumerate}
Theorem 4 can be used backwards recursively to obtain the asymptotic distribution of the parameters at each stage since the oracle properties can be inherited from the prior iteration. The plug-in variance formula (\ref{eq-var-2}) can also be used backwards recursively for inference for the parameters in each stage.

\section{Simulation Studies}
\cite{Cha09}  designed a thorough simulation study of two-stage
Q-learning, which covers regular, non-regular and close-to-non-regular
conditions. In this section, we apply the proposed method to the
same simulation study conditions. Specifically, a total of $n=300$
subjects are contained in the data. We set $R_1=0$ and  $(O_1,
A_1, O_2, A_2, R_2)$ is collected on each subject. The binary
covariates $O_t$'s and the binary treatments $A_t$'s are generated
as follows:
\[
P(O_1=1)=P(O_1=-1)=1/2,
\]
\[P(A_t=1)=P(A_t=-1)=1/2,t=1,2,\]
\[
P(O_2=1|O_1,A_1) =
1-P(O_2=-1|O_1,A_1)=expit(\delta_1O_1+\delta_2A_1),
\]
where $expit(x)=\exp(x)/(1+\exp(x))$.
\[
R_2=\gamma_1+\gamma_2O_1+\gamma_3A_1+\gamma_4O_1A_1+\gamma_5A_2+\gamma_6O_2A_2+\gamma_7A_1A_2
+ \varepsilon,
\]
where $\varepsilon \sim N(0,1)$. The Q-functions for time $t=1,2$
are both correctly specified as:
\begin{align}
Q_2(O_1,A_1,O_2,A_2;\bbeta_2,\bpsi_2)= &
\beta_{21}+\beta_{22}O_1+\beta_{23}A_1+\beta_{24}O_1A_1 \nonumber \\
& + \psi_{21}A_2+\psi_{22}O_2A_2+\psi_{23}A_1A_2,\\
Q_1(O_1,A_1;\bbeta_1,\bpsi_1)=&\beta_{11}+\beta_{12}O_1 +
\psi_{11}A_1 + \psi_{12}O_1A_1.
\end{align}
As shown in
\cite{Cha09}, the true values of $\bpsi_1 \equiv (\psi_{110}, \psi_{120})^T$ are
given by:
\[
\psi_{110}=\gamma_3 + q_1|f_1|- q_2|f_2|+(1/2-q_1)|f_3|
-(1/2-q_2)|f_4|,
\]
\[
\psi_{120}=\gamma_4+q_1'|f_1|- q_2'|f_2|-q_1'|f_3| +q_2'|f_4|,
\]
%\begin{align*}
%\psi_{11}&= \gamma_4 + (p(1,1)-p(-1,1))|f_1| + (q(1,1)-q(-1,1))|f_3| +
%|f_2| (p(-1,-1)-p(1,-1)) + |f_4|(q(-1,-1)-q(1,-1))\\
%&= \gamma_4 + q_1'|f_1| -q_3'|f_3| -q_2' |f_2| + q_4'|f_4|\\
%\end{align*}
where $q_1=1/4(expit(\delta_1+\delta_2) +
expit(-\delta_1+\delta_2))$, $q_2=1/4(expit(\delta_1-\delta_2) +
expit(-\delta_1-\delta_2))$, $q_1'=1/4(expit(\delta_1+\delta_2) -
expit(-\delta_1+\delta_2))$, $q_2'=1/4(expit(\delta_1-\delta_2) -
expit(-\delta_1-\delta_2))$, $f_1=\gamma_5+\gamma_6+\gamma_7$,
$f_2=\gamma_5+\gamma_6-\gamma_7$,
$f_3=\gamma_5-\gamma_6+\gamma_7$,
$f_4=\gamma_5-\gamma_6-\gamma_7$. Let
$\bgamma=(\gamma_1,...,\gamma_7)^T$. We consider the following six settings:
\begin{description}
\item[Setting 1:] $\bgamma=(0,0,0,0,0,0,0)^T,\delta_1=\delta_2=0.5$.
\item[Setting 2:] $\gamma= (0,0,0,0,0.01,0,0)^T,\delta_1=\delta_2=0.5$.
\item[Setting 3:] $\gamma= (0,0,-0.5,0,0.5,0,0.5)^T,\delta_1=\delta_2=0.5$.
\item[Setting 4:] $\gamma= (0,0,-0.5,0,0.5,0,0.49)^T,\delta_1=\delta_2=0.5$.
\item[Setting 5:] $\gamma= (0,0,-0.5,0,1,0.5,0.5)^T,\delta_1=1, \delta_2=0$.
\item[Setting 6:] $\gamma= (0,0,-0.5,0,0.25,0.5,0.5)^T, \delta_1=\delta_2=0.1$.
\end{description}
The values of $\bpsi_{20}^T\bS_{22}$ for each setting are listed in Table \ref{tab:uniq}. % As shown in the table, the examples 1, 3 and 5 are non-regular since some values of $S_{21}^T\bpsi_2^*$ are exactly equal to zero. The examples 2, 4 and 6 are regular settings, among which example 2 is close to the non-regular setting in example 1, example 4 is close to example 3 while example 6 is a full regular setting.

\begin{table}
\caption{Values of $\bpsi_{20}^T\bS_{22}$ in the simulation studies.}
\centering
\fbox{%
\begin{tabular}{ccccccc}
%\hline
 & \multicolumn{4}{c}{$\bS_{22}=(1,O_2,A_1)$}\\
Setting& (1,1,1) &(1,1,-1) & (1,-1,1) & (1,-1,-1)\\
\hline
1 & 0 & 0& 0 & 0\\
2 & 0.01 & 0.01 & 0.01 & 0.01\\
3 & 1 & 0 & 1 & 0\\
4 & 0.99 & 0.01 & 0.99 & 0.01\\
5 & 2 & 1 & 1 & 0\\
6 & 1.25 & 0.25 & 0.25 & -0.75\\
\end{tabular}}
\end{table}\label{tab:uniq}

We applied the proposed penalized Q-learning with the adaptive lasso penalized regression to the six settings.  The one-step LQA algorithm is used with least squares estimation for the initial values. The tuning parameter $\lambda$ in the
adaptive lasso penalty is chosen by five fold cross-validation.
When the estimated value $|\widehat\bpsi_2^T\bS_{22}|<0.001$, it will
be set as zero in the stage-1 estimation.  The simulation results
shown in Table \ref{tab:sim} were summarized over 1000
replications. We included both the oracle estimator and the hard-max estimator
for comparison. Theoretical standard errors and 95\% confidence
intervals for the hard-max estimator are not available.

Setting 1 is a completely non-regular setting, where $\bpsi_{20}^T\bS_{22}=0$ for all values of $\bS_{22}$. The oracle estimator automatically sets  $\hbpsi_{2}=0$ and is therefore regular. It has a very small bias, with standard errors accurately predicted by the theory and  95\% confidence interval coverage close to the nominal value.  The PQ estimator's performance is very close to the oracle estimator, with similar bias, a slightly bigger but still well estimated standard error and similar confidence intervals. %HM, HT$_{0.08}$, HT$_{0.20}$ all have ignorable bias (Table \ref{tab:cha09} but the percentile bootstrapped confidence intervals show significant overcoverage, indicating non-regular estimation. The soft threshold estimator has a similar performance as the oracle and PQ estimators.

Setting 2 is regular but very close to setting 1 with $\bpsi_{20}^T\bS_{22}$ all equal to 0.01. The oracle estimator reduces to the hard-max. Although its bias is small, the oracle estimator's  theoretical standard error is significantly overestimated and, as a result, its confidence intervals show significant over-coverage for $\psi_{110}$. On the other hand, the PQ estimator remains consistent and its standard error estimation remains close to the empirical values. %The performances of HM, HT$_{0.08}$, HT$_{0.20}$ and ST are all similar to example 1.

Setting 3 is another non-regular setting. The value of
$\bpsi_{20}^T\bS_{22}$ is equal to 1 for half of the subjects and
equal to 0 for the other half. As expected, the
oracle estimator has the best performance in the sense of having smallest
bias and standard error as well as being precisely predicted by the theoretical standard error. The PQ estimator has a bigger bias and standard error than the oracle estimator but the theoretical standard error remains close to the empirical values.

Setting 4 is a regular setting but close to Setting 3. The PQ
estimator outperforms the oracle estimator, with both a smaller
bias and a smaller standard error. This phenomena is consistent
with findings in Setting 2, which is another nearly non-regular setting.
However, in Setting 2, we did not observe substantial overestimation of the standard
error nor over-coverage of the confidence intervals in the oracle
estimator.
%which is equal to hard-max estimator, has a significant bias and an overestimated standard error. Although its 95\% CI coverage probability is perfect, this results from a mixed effect of undercoverage due to bias and overcoverage due to overestimated standard error and should not be considered as a sign of regularity. In contrast, PQ has small bias, consistent standard error estimation and coverage probability.

Setting 5 is a non-regular setting similar to Setting 3 and the
performances of the oracle and PQ estimators are very similar to their performances in
Setting 3.

Setting 6 is a completely regular setting with values of
$\bpsi_{20}^T\bS_{22}$ well above zero. The PQ estimator has a
slightly bigger bias and a slightly bigger standard error than
the oracle estimator. Both estimators's standard errors are well predicted from
the theory.

In summary, the behavior of the PQ estimator, including its bias, theoretical computed standard error and coverage probability of theoretically computed 95\% confidence
intervals, are consistent in all six settings, whether regular or
non-regular. In  non-regular or completely regular settings, PQ-learning
usually has bigger bias and larger standard error than the oracle
estimator. In  regular settings which are close to non-regular
cases, it has smaller bias and standard error than the oracle
estimator.

\begin{table}
\caption{Summary statistics and empirical coverage probability of 95\% nominal percentile CIs for $\psi_{110}$ and $\psi_{120}$ using the oracle estimator, the proposed PQ-learning based (PQ) estimator and the hard max (HM) estimator. Specifically, ``Std-MC'' refers to the standard deviation of 1000 estimates for $\psi_{110}$ or $\psi_{120}$,
``Std'' refers to the average of the 1000 standard error  estimates and ``CP'' refers to the empirical coverage probability of 95\% nominal percentile confidence interval. A ``*'' indicates a significantly different coverage rate other than the nominal rate.}
\centering
\fbox{%
\begin{tabular}{clrrrrrrrr}
%\hline
 & & \multicolumn{4}{c}{$\psi_{110}$} & \multicolumn{4}{c}{$\psi_{120}$}\\
Setting &  &  Bias & Std-MC & Std & CP & Bias & Std-MC & Std & CP\\
\hline
1 & oracle & -0.0015 & 0.058 & 0.058 & 94.6 & -0.0004 & 0.060 & 0.058 & 94.3 \\
   & PQ & -0.0013 & 0.060 & 0.061 & 95.1 & -0.0009 & 0.060 & 0.058 & 94.0 \\
   & HM & -0.0005 & 0.066 &$-$  &$-$  & -0.0023 & 0.060 & $-$ &  $-$\\
\\
2 & oracle & -0.0025 & 0.065 & 0.075 & 97.5$^*$ & 0.0003 & 0.056 & 0.059 & 96.3 \\
   & PQ & -0.0026 & 0.059 & 0.060 & 94.0 & 0.0013 & 0.056 & 0.058 & 96.3 \\
   & HM & -0.0025 & 0.065 &$-$  &$-$   & 0.0003 & 0.056 & $-$  & $-$  \\
\\
3 & oracle &-0.0032 & 0.071 & 0.071 & 95.3 & -0.0043 & 0.057 & 0.058 & 94.9 \\
   & PQ &-0.0182 & 0.073 & 0.076 & 95.2 & -0.0049 & 0.057 & 0.058 & 95.4 \\
   & HM & -0.0437 & 0.075 & $-$  &  $-$& -0.0051 & 0.058 & $-$ & $-$ \\
\\
4 & oracle & -0.0330 & 0.076 & 0.079 & 94.9 & -0.0021 & 0.058 & 0.059 & 94.9 \\
   & PQ & -0.0073 & 0.074 & 0.075 & 95.5 & -0.0022 & 0.058 & 0.058 & 94.8 \\
   & HM &  -0.0330 & 0.076 &  $-$&$-$  & -0.0021 & 0.058 & $-$ &$-$  \\
\\
5 & oracle & -0.0002 & 0.075 & 0.079 & 95.7 & 0.0005 & 0.067 & 0.063 & 93.9 \\
& PQ &  -0.0188 & 0.079 & 0.080 & 95.3 & 0.0056 & 0.067 & 0.064 & 94.0 \\
&HM &  -0.0204 & 0.077 & $-$ &$-$  & 0.0092 & 0.067 & $-$ & $-$ \\
\\
6 &oracle &  0.0003 & 0.078 & 0.080 & 95.4 & 0.0009 & 0.063 & 0.062 & 94.8 \\
   & PQ & -0.0012 & 0.079 & 0.080 & 95.3 & 0.0009 & 0.063 & 0.062 & 94.7 \\
   & HM & 0.0003 & 0.078 & $-$ &$-$  & 0.0009 & 0.063 & $-$ &  $-$\\
\end{tabular}}
\end{table}\label{tab:sim}

\cite{Cha09} proposed several bootstrapped confidence intervals (PB: percentile bootstrap, HB: hybrid bootstrap, DB: double bootstrap) for the hard-max estimator as well as hard-thresholding
estimators with threshold at 0.08 (HT$_{0.08}$) and 0.20
(HT$_{0.20}$) and soft-thresholding estimator (ST). We include
their simulation results ($\psi_{110}$ only) in Table \ref{tab:cha09} for comparison. The coverage probabilities of all 10 inferential methods in the six settings are plotted in Figure 2. The boxplot of oracle estimator, PQ-estimator and hard-max estimator of $\psi_{11}$ and $\psi_{12}$ from 1000 estimates of these six settings are provided in Figures 3 and 4 respectively. The PQ-estimator has smaller bias and narrower inter-quantile range compared with the hard-max estimator in most of the settings. The bias and Monte-Carlo standard deviation of the hard-max estimator presented in Table 2 and Table 3 are similar, validating a direct comparison
between the two study results. Readers are directed to the
original articles for a discussion on the performances of different
estimators and bootstrapping methods.

Briefly speaking, the bias of the hard-max estimator in Settings 3 and 4 is relatively big.
%the hard-max estimator with the computationally intensive double bootstrapped confidence interval show correct coverage rates in all six settings.
 The percentile bootstrapped and hybrid bootstrapped
confidence intervals can correct the coverage rates for
hard-thresholding and soft-thresholding estimators in some
settings. However, neither of these two bootstrapped methods can
consistently improve the coverage rate for all estimators. Overall,
the soft-thresholding estimator has the best performance with the
percentile bootstrapped confidence intervals. This is consistent
with our findings for the PQ estimator due to their similar nature.
Nonetheless, we derived the theoretical formula for standard errors
and therefore did not need to rely on bootstrap method. Thus PQ-learning is
substantially more computationally efficient than the soft-thresholding approach
with the bootstrap and performs at least as well.

\begin{table}\label{tab:cha09}
\caption{Summary statistics and empirical coverage probability of 95\% nominal percentile CIs for $\psi_{110}$ and $\psi_{120}$ using the hard max (HM) estimator, the hard threshold estimator with $\alpha=0.08$ $(HT_{0.08})$ and $\alpha=0.20$ $(HT_{0.20})$, and the soft-threshold estimator quoting the simulation results from \cite{Cha09}. Specifically, ``Var'' denotes the sample variance of 1000 estimates for $\psi_{110}$ or $\psi_{120}$. ``PB'', ``HB'' and ``DB'' denote percentile bootstrap, hybrid bootstrap and double bootstrap, respectively. ``CP'' and ``*'' have the same meaning as given in Table 2.}
\centering
\fbox{
\begin{tabular}{clrrlll}
Setting & Estimator &  Bias & Var & \multicolumn{3}{c}{CP} \\
 & & & & PB & HB & DB\\
\hline
1 & HM & 0.0003 & 0.0045 & 96.8$^*$ & 93.5$^*$ & 93.6\\
& HT$_{0.08}$ & 0.0017 & 0.0044 & 97.0$^*$ & 95.0 & $-$\\
& HT$_{0.20}$ & 0.0002 & 0.0050 & 97.4$^*$ & 92.8$^*$ & $-$\\
&ST & 0.0009 & 0.0036 & 95.3 & 96.1 & $-$\\
2 & HM & 0.0003 & 0.0045 & 96.7$^*$ & 93.4$^*$ & 93.6\\
& HT$_{0.08}$ & 0.0010 & 0.0044 & 97.1$^*$ & 95.3&$-$\\
& HT$_{0.20}$ & 0.0003 & 0.0050 & 97.3$^*$ & 93.5$^*$ & $-$\\
& ST & 0.0008 & 0.0036 & 95.4 & 95.9 & $-$\\

3 & HM &0.0401& 0.0059 & 88.4$^*$ &92.7$^*$ &94.8\\
& HT$_{0.08}$&0.0083& 0.0058& 94.3& 94.3 & $-$\\
&HT$_{0.20}$& 0.0179& 0.0062& 93.5$^*$ &93.5$^*$&$-$\\
&ST& 0.0185& 0.0055 & 93.4$^*$& 94.9&$-$\\

4 & HM &0.0353& 0.0059& 89.6$^*$ &93.1$^*$ &94.4 \\
&HT$_{0.08}$& 0.0037& 0.0058 & 94.6& 94.1&$-$\\
&HT$_{0.20}$& 0.0130& 0.0062& 93.9& 92.8$^*$&$-$\\
&ST& 0.0138& 0.0055& 94.1& 95.0& $-$\\

5 & HM& 0.0209& 0.0069& 92.7$^*$ &93.1$^*$ &94.2 \\
&HT$_{0.08}$& 0.0059 &0.0070&  93.9 &93.2$^*$ &$-$\\
&HT$_{0.20}$& 0.0101& 0.0072 & 93.3$^*$ &93.0$^*$&$-$  \\
&ST& 0.0065& 0.0069& 93.8& 94.6 & $-$\\

6 & HM & 0.0009 &0.0067& 95.0& 93.8 &95.0\\
&HT$_{0.08}$& 0.0003& 0.0081& 95.1& 88.5$^*$&$-$ \\
&HT$_{0.20}$& 0.0011& 0.0074 & 94.8 &91.2$^*$&  $-$\\
&ST& 0.0052 &0.0074& 94.8& 91.7$^*$& $-$\\
\end{tabular}}
\end{table}

\begin{figure}[htbp]
\centering
%\hspace{0.033\textwidth}
\rotatebox{270}{
    \makebox{\includegraphics[width=6in, height=6.5in]{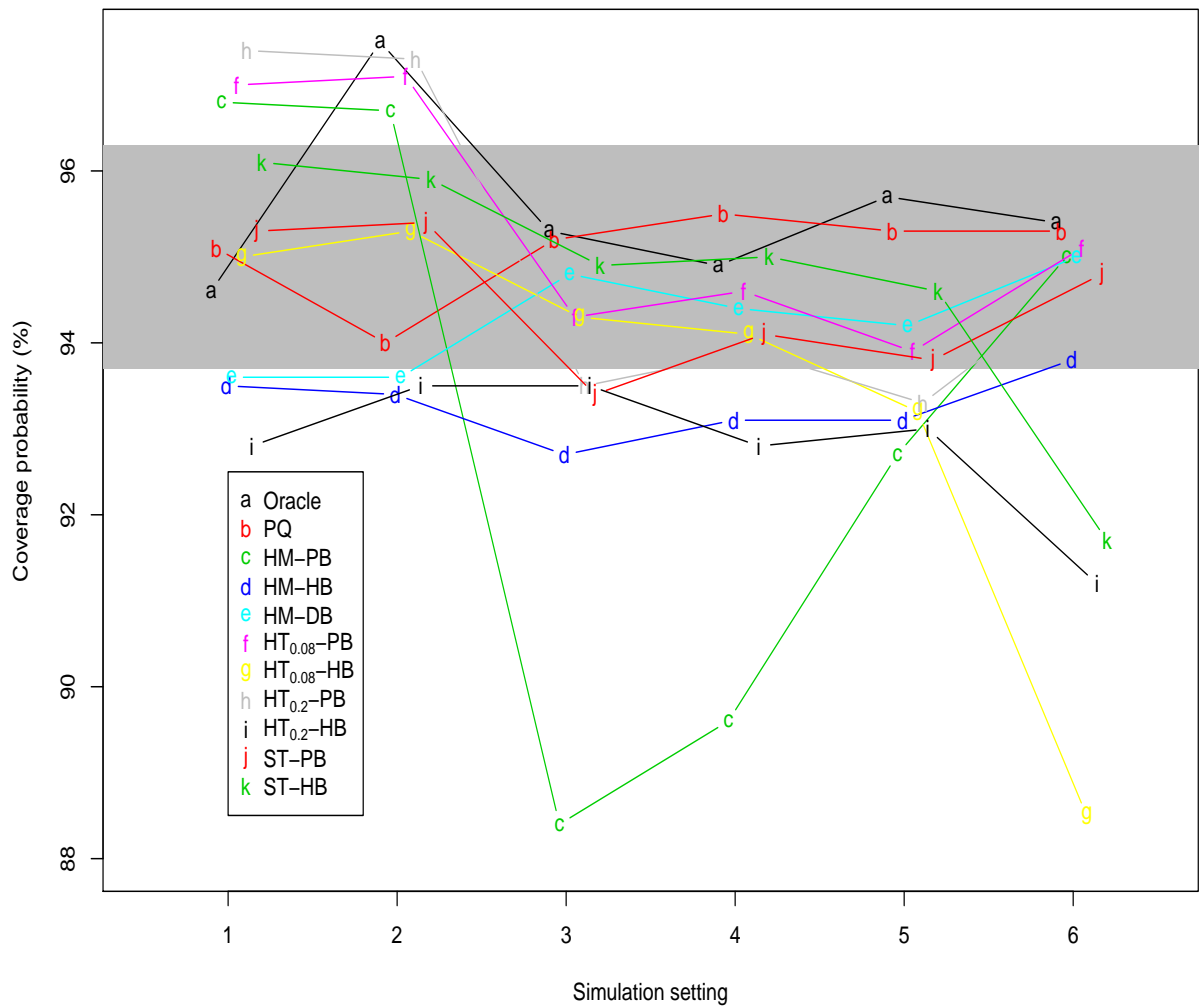}}}
   \caption{Plot of coverage probabilities with all ten inference methods in six settings, where the shaded area is 95\% confidence for monte-carlo error: [93.7\%, 96.3\%].}
\label{f3}
% \hspace{0.033\textwidth}
%\hfill
\end{figure}

\begin{figure}[htbp]
\centering
%\hspace{0.033\textwidth}
\rotatebox{270}{
    \makebox{\includegraphics[width=7in, height=7.2in]{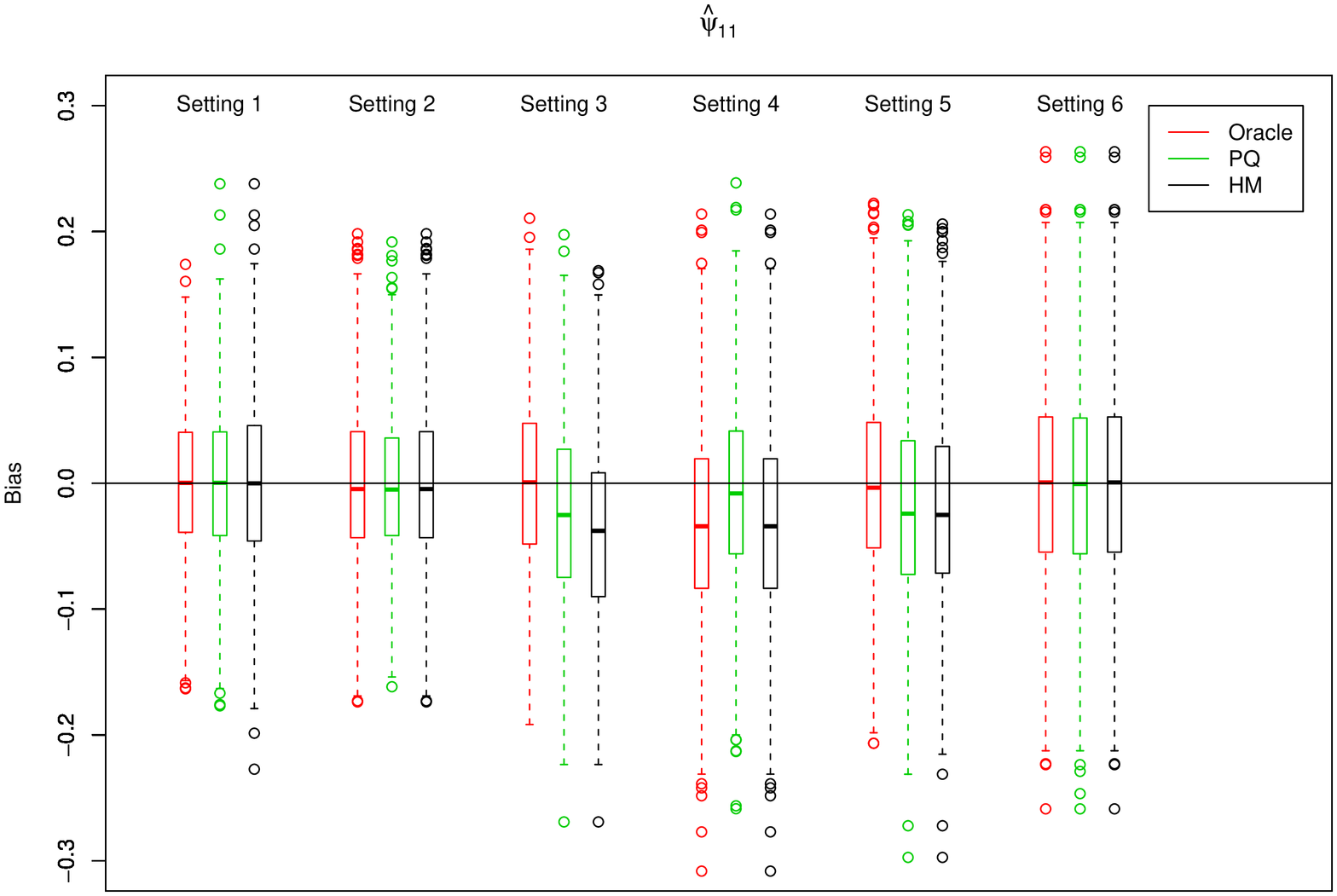}}}
   \caption{The boxplot of oracle estimator, PQ-estimator and hard-max estimator of $\psi_{11}$ from 1000 estimates of the six settings.}
\label{f4}
% \hspace{0.033\textwidth}
%\hfill
\end{figure}

\begin{figure}[htbp]
\centering
%\hspace{0.033\textwidth}
\rotatebox{270}{
    \makebox{\includegraphics[width=7in, height=7.2in]{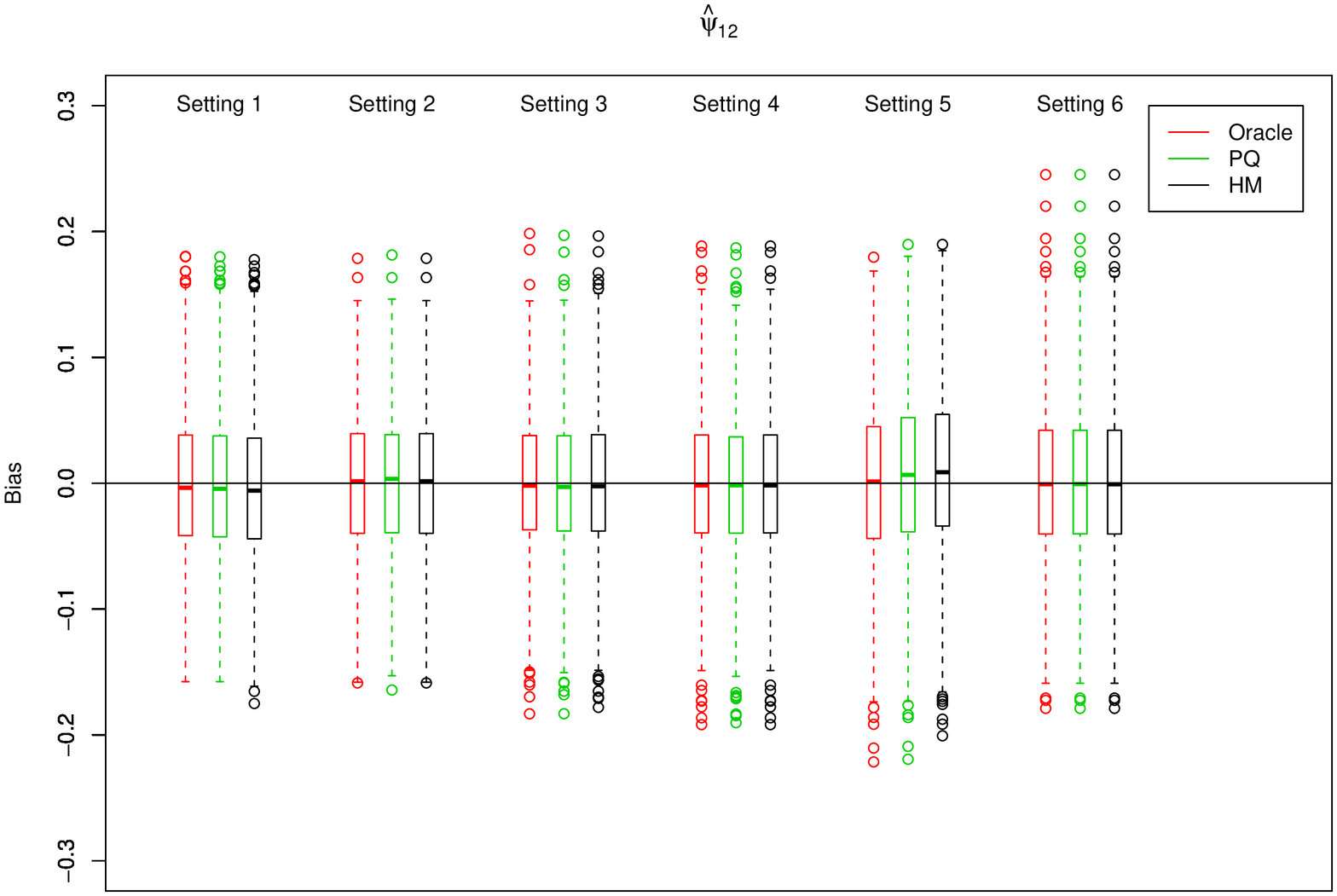}}}
   \caption{The boxplot of oracle estimator, PQ-estimator and hard-max estimator of $\psi_{12}$ from 1000 estimates of the six settings.}
\label{f4}
% \hspace{0.033\textwidth}
%\hfill
\end{figure}

\section{Analysis of STAR*D study}
We here present the analysis of STAR*D study data using
PQ-learning. STAR*D is a prospective multi-site
randomized clinical trial designed to determine the comparative
effectiveness of different multi-level treatment options for patients
with major depressive disorder (MDD). A total of 4041 patients (ages
18-75) with nonpsychotic MDD were enrolled and initially treated with
citalopram (CIT) (Level 1 treatment) for a minimum of 8
weeks, with strong encouragement to complete 12 weeks in order to
maximize benefit. During this and all subsequent treatment levels,
patients would have clinic visits at weeks
0, 2, 4, 6, 9 and 12.

During all clinic visits, symptomatic status would
be measured by the 16-item Quick Inventory of Depressive
Symptomatology ¡V Clinician-Rated (QIDS-C$_{16}$).
Patients who did not have a satisfactory response to treatments, defined
as either $< 50\%$ reduction in QIDS-C$_{16}$ or QIDS-C$_{16}>5$,
would be elligible for Level 2 treatment. Seven treatment options are available
at Level 2, which can be classified into two classes, (1) Medication or
Psychotherapy Switch: sertraline (SER), venlafaxine (VEN), bupropion (BUP) or Cognitive
Psychotherapy (CT); and (2) Medication or Psychotherapy Augmentation:
 CIT+BUP, CIT+buspirone (BUS) or CIT+CT. Patients who were assigned to
 CT or CIT+CT  in Level 2 and did not have a satisfactory response
 would be elligible for Level 2A, during which they would be  treated
 with either VEN  or BUP.
Patients who did not respond
 satisfactorily at Level 2 and Level 2A, if applicable, would continue
 to Level 3  treatment. Level 3 includes four options: Medical Switch
 to mirtazapine (MIRT) or nortriptyline (NTP), and Medical augmentation
  with either  lithium or  thyroid hormone added to level 2 or 2A
  treatments. Patients who did not respond satisfactorily to Level 3
  treatments would continue to Level 4 treatments, which include two
  options: switch to tranylcypromine or MIRT+VEN. For a complete
  description of STAR*D, see \cite{ISI:000182929300010} and \cite{ISI:000220636100012}.

In this analysis, for demonstration purpose, we consider a subgroup of  STAR*D patients, the
112 patients who were randomized to either BUP or SER in Level 2, had
no satisfactory response at the end of Level 2, and were then
randomized to either MIRT or NTP in Level 3.  The analysis focuses
on selecting the optimized treatment regime at Level 2 and Level 3,
out of the 4 unique treatment combinations. Since the higher
QIDS-C$_{16}$ is, the more severe the symptom is, we define the reward
as negative of QIDS-C$_{16}$ collected at the end
of the Level 3. Similarly as discussed in \cite{ISI:000246387600007}, the state
variable used  to tailor individual treatment is the changing rate of
QIDS-C$_{16}$ during the previous treatment level. We dichotomize the changing rates at
zero. Two patients were further removed due to  missing values in  the reward or the tailor
variables.

Following the
notations in the simulation study, let $O_1$ and $O_2$ be the indicator of whether the QIDS-C$_{16}$
changing rate  is greater than zero in Level 1 and Level 2
respectively. Let $A_1=1$ if Level 2
treatment is SER and $A_1=-1$ if it is BUP. Let $A_2=1$ if Level 3
treatment is  NTP and $A_2=-1$ if it is MIRT, $R_2 = -QIDS-C_{16}$
collected at the end
of  Level 3.
The Level 3 regression model is:
\[
R_2=\beta_{21}+\beta_{22}O_1+\beta_{23}A_1 + \beta_{24}O_2 + \psi_{21}A_2+\psi_{22}A_2O_1+\psi_{23}A_2A_1 + \psi_{24}A_2O_2+\varepsilon_2.
\]
Since  the main effects of $A_1$ and $O_2$ are not
statistically significant, we did not include additional  interaction
terms in Level 3 model.

\begin{table}
\caption{\label{tab:real1}Level 3 regression model coefficients estimation using both unpenalized least square estimation and individual penalized least square estimation.}
\centering
\fbox{
\begin{tabular}{lrlrl}
%\hline
Variable & \multicolumn{2}{c}{unpenalized} &
\multicolumn{2}{c}{penalized}\\
 & Coefficient & 95\% CI & Coefficient & 95\% CI\\
\hline
Intercept & -13.165 & (-14.349, -11.981) & -13.185 & (-14.330, -12.039)\\
  $O_1$ & -1.202 & (-2.348, -0.057) & -1.124 & (-2.233, -0.015) \\
  $A_1$ & 0.004 & (-0.945, 0.954) & -0.046 & (-0.967, 0.874) \\
  $O_2$ & -0.587 & (-1.605, 0.431) & -0.554 & (-1.533, 0.425) \\
  $A_2$ & -1.276 & (-2.460, -0.092) & -1.266 & (-2.239, -0.292) \\
  $O_1A_2$ & -1.621 & (-2.766, -0.475) & -1.300 & (-2.410, -0.191) \\
  $A_1A_2$ & 0.535 & (-0.414, 1.484) & 0.052 & (-0.775, 0.880) \\
  $O_2A_2$ & 0.278 & (-0.740, 1.297) & 0.017 & (-0.748, 0.783) \\
%\hline
\end{tabular}}
\end{table}

Table 4 shows the Level 3 regression model coefficients
estimation using both unpenalized standard least square estimation and individual
penalized least square estimation. Qualitatively, unpenalized and penalized estimations
are consistent. Patients whose symptom worsened (i.e., $O_1=1$ or QIDS-C$_{16}$
increased) during Level 1 would have worse reward. Level 2 treatments (SER versus BUP) as well as QIDS-C$_{16}$
changing rate during Level 2 show no differential effect on the
final outcome. However, the two Level 3 treatment options show significant different effects on
patients with $O_1=1$ versus patients with $O_1=-1$. Among patients
whose symptom worsened in Level 1,  NTP further worsened their
symptom when compared to MIRT. Among patients whose symptom improved
in Level 1, NTP and MIRT show no
obvious difference as to the final outcome.

Quantitatively, the penalized estimator
has smaller standard errors in the coefficient estimation of
$\bpsi_2=(\psi_{21},\psi_{22},\psi_{23},\psi_{24})^T$ than the
unpenalized estimator. In addition, the penalized estimator
dramatically shrinks
coefficients of the two unimportant predictors $A_1A_2$ and $O_2A_2$
toward zero. On the other hand, these two
estimators are similar in the coefficient estimation of
$\bbeta_2=(\beta_{21},\beta_{22},\beta_{23},\beta_{24})^T$, which is
expected since the penalty is imposed only on $\bpsi_2$.
In order to shrink  coefficients of the unimportant
predictors  $A_1$ and $O_2$, one can further
impose penalty on $|\bbeta_2|$, which will not be implemented in
this work. The lack of effect of $A_1$ and $O_2$ is actually expected
since we include in this analysis only patients eligible for Level 3
treatment, in other words, only patients who did not respond
satisfactorily to Level 2 treatment. This inclusion criteria is
imposed because our current framework is built on the
situation where all patients will be treated in both stages. The
extension to cases where patients may be cured during intermediate
stages and hence not eligible for subsequent treatment stages is
not trivial and will be considered in future work.

\begin{table}\label{tab:real2}
\caption{Values of $|\bS_{22}^T\widehat\bpsi_{2}|$ in STAR*D study.}
\centering
\fbox{
\begin{tabular}{rrrrrr}
 % \hline
 & &&& \multicolumn{2}{c}{$|\bS_{22}^T\widehat\bpsi_{2}|$ }\\
&O1 & A1 & O2 & Unpenalized & Penalized\\
  \hline
 & -1 & -1 & -1 & 0.468 & 0.035 \\
   & -1 & -1 & 1 & 0.088 & 0.000 \\
   & -1 & 1 & -1 & 0.601 & 0.070 \\
   & -1 & 1 & 1 & 1.158 & 0.105 \\
   & 1 & -1 & 1 & 3.153 & 2.601 \\
   & 1 & 1 & -1 & 2.640 & 2.531 \\
   & 1 & -1 & -1 & 3.710 & 2.636 \\
   & 1 & 1 & 1 & 2.083 & 2.496 \\
\end{tabular}}
\end{table}

Table \ref{tab:real2} shows the estimated values of
$|\bS_{22}^T\bpsi_{2}|$, where $\bS_{22}=(1,O_1,A_1,O_2)^T$.
When $O_1=-1$, Level 3 treatment effect is small but
 the unpenalized estimator shows  significant bias from
zero. On the other hand, the penalized estimator successfully shrinks
the value of $|\bS_{22}^T\bpsi_{2}|$ in all groups close to zero.
Although due to the limitation of current LQA algorithm, the penalized estimator
cannot exactly set $|\bS_{22}^T\bpsi_{2}|$ to zero, the bias is
significantly smaller than unpenalized estimator. When $(O_1,A_1,O_2)=(-1,-1,1)$, the penalized estimation of $|\bS_{22}^T\bpsi_{2}|$
falls below the preselected cutoff of 0.001 and is shown as 0 in Table \ref{tab:real2}.
When $O_1=1$, the treatment option MIRT can significantly improve the
symptom. Since $A_1$ and $O_2$ have no important effect on the
outcome, we expect similar treatment effects among the four groups
with $O_1=1$. From this point of view, the
unpenalized estimator is inferior since it shows much bigger variation than the penalized
estimator.

\begin{table}
\caption{Level 2 regression model coefficients
estimation using both Hard-max and PQ-learning.}
\centering
\fbox{
\begin{tabular}{rrlrl}
 & \multicolumn{2}{c}{Hard-Max} & \multicolumn{2}{c}{PQ-learning}\\
Variable& Coefficient &Hybrid 95\% CI & Coefficient & 95\% CI\\
  \hline
Intercept & -11.063 & (-12.482, -10.095) &-11.612 & (-13.076, -10.149) \\
  $O_1$ & 0.263 &(-0.764, 1.547) &0.313 & (-1.114, 1.740) \\
  $A_1$ & -0.119 & (-1.120, 0.884) &-0.038 & (-1.115, 1.039) \\
  $O_1A_1$ & -0.448 & (-1.079, 0.251) &-0.085 & (-0.830, 0.661) \\
\end{tabular}}
\end{table}\label{tab:real3}

We next consider the Level 2 regression model. The pseudo-outcome
$\widehat{Y}$ is defined by
$
\widehat{Y}=\bbeta_2^T\bS_{21} + |\bpsi_2^T\bS_{22}|
$ and we impose the following Level 2 model
\[
\widehat{Y}=\beta_{11}+\beta_{12}O_1+\beta_{13}A_1 + \beta_{14}O_1A_1.
\]
Table 6 shows the Level 2 model coefficient
estimation using both Hard-max and PQ-learning. The coeffiecients estimations for
the intercept and $O_1$ are similar from two different estimation
methods. While in the estimation of cofficients for $A_1$ and
$O_1A_1$, PQ-learning's estimation is significantly towards
zero. Based on 95\% CI from PQ-learning, $O_1$, $A_1$ have no effect
on the pseudo-outcome. Since $A_1$ shows no effect in Level 3
regression either, it is easy to interpret its lack of effect on
pseudo-outcome. In contrast, $O_1$ is a strong predictor in Level 3
treatment. Its lack of effect in Level 2 regression may be explained
as follows. In Level 3 regression, the $O_1=1$ group's reward is
smaller than the  $O_1=-1$ group's reward  by
$2\beta_{22} \approx 2.24$. However, the optimal Level 3 treatment can
increase the $O_1=1$ group's reward  by $|\bpsi_2^T\bS_{22}| \approx
2.6$ but cannot increase the $O_1=-1$'s reward. Combined together,
$O_1$ has no effect on the pseudo-outcome.

Our analysis found that the optimal Level 2 and Level 3 treatment
regime is following. If a patient's  symptom worses in Level 1 and
remains unsatisfactory in Level 2, MIRT is a better option for Level 3
treatment when compared to  NTP. If a patient's symptom improves in Level 1
and remains unsatisfactory in Level 2, MIRT or NTP have similar effect
as Level 3 treatment.

\section{Discussion}
In this article, we propose a penalized Q-learning framework and an individual selection procedure for developing optimal dynamic treatment regimes. Statistical inference for parameters at each stage are established. The long-term difficulty in developing optimal dynamic treatment regimes---non-regularity associated with the treatment effect parameters---is solved. The methods are shown to be effective and the standard errors are estimated computationally efficiently and with good accuracy.

The proposed concept of individual selection is generally applicable. Specifically, the Q-learning approach is an inefficient special case of the doubly robust structural nested mean model (drSNMM) proposed by Robins (2004). The drSNMM is an estimating equation approach, which also has the difficulty of non-regularity. The PQ-learning approach proposed here can be straightforwardly extended to penalized drSNMM to handle the non-regularity issue.

Although the linear model form of the Q-functions presented here is an important first step, as well as being useful for illustrating the ideas of this paper, this form may not be sufficiently flexible for certain practical settings. Semiparametric models are a potentially very useful alternative in many such settings because such models involve both a parametric component which is usually easy to interpret and a nonparametric component which allows greater flexibility. Generalizations of Q-functions to allow diverse data such as ordinal outcome, censored outcome and semiparametric modeling, are thus future research topics of practical importance.

The current theoretical framework is based on discrete covariates. This condition is not as restricted as it looks. For example, in a practical two-stage setting where continuous covariates are presented, unless in the rare case where the parameter $\bpsi_{20}$ is zero, the ``problematic'' set $\mathcal{M}^c_{\star} = \{i: \bpsi_{20}^TS_{22i}=0 \}$ will not have positive probability. Having said that, we can always discretize the continuous covariates into several strata and apply the proposed methods with these strata. Obviously, there will be loss of information with this approach. Future research to extend our work to continuous covariates would also be very useful in practice. Likewise, the current PQ-learning framework works for two-level treatments. The generalization to multilevel treatments will be a natural and useful next step.

%For the individual selection component, we have shown that many popular penalty functions are applicable theoretically, including the SCAD and adaptive lasso. In this paper, we have only applied the adaptive lasso for demonstration. It would be interesting to study more penalty functions and to develop more efficient algorithms for individual selection in future work.

In many clinical studies, the state space is often of very high dimension. To develop optimal dynamic treatment regimes in this case, it will be important to develop simultaneous variable selection (for state variables) and individual selection. More modern machine learning techniques such as support vector regression and random forests can be nested into our PQ-learning framework as powerful tools to develop optimal dynamic treatment regimes.
%$p_{\lambda_n}(|Q_2(\bS_2,1;\btheta_2)-Q_2(\bS_2,0;\btheta_2)|).$

%Consider high-dimensional state space. Sparsity (state variable selection).

%\section*{Acknowledgements}

\bibliographystyle{ims}
\bibliography{prop-ref}
\end{document}